\documentstyle[12pt,cite,amssymb,amsmath]{article}
\def\cf#1#2#3#4{\bibitem{#1}{#2}.~{\it #3}\,;~{#4}.}

\def\cy{Calabi--Yau~}
\def\C{{\mathbb{C}}}

\def\eq#1{(\ref{#1})}
\def\eqs#1#2{(\ref{#1})--(\ref{#2})}
\def\F{{\mathcal{F}}}
\def\FI{Fayet--Iliopoulos~}
\def\G{\tilde{{\mathcal{G}}}}
\def\goth#1{{\mathfrak #1}}

\def\ie{{\em i.e.~}}
\def\ker#1{\qopname\relax o{Ker}~{#1}}
\def\p{{\mathbb{P}}}
\def\pa{\partial}
\def\re{\goth{Re}}
\def\R{{\mathbb{R}}}
\def\rt{\longrightarrow}

\def\tilde{\widetilde}

\def\T{{\mathbb{T}}}

\def\Z{{\mathbb{Z}}}
\def\alg#1{alg-geom/{#1}}
\def\dgga#1{dg-ga/{#1}}
\def\hep#1{hep-th/{#1}}
\def\jdg#1#2#3{J. Diff. Geo. {\bf {#1}}~({#2})~{#3}}
\def\npb#1#2#3{Nucl.~Phys. {\bf B{#1}}~({#2})~{#3}}
\def\plb#1#2#3{Phys. Lett. {\bf B{#1}}~({#2})~{#3}}

\begin{document}
\title{\vspace{-3cm}\hfill
\parbox[t]{2in}{\centering{\small ROM2F-98/10 \\ hep-th/9803192}}\\  
\vspace{4cm}\centerline{A Ricci-flat metric on D-brane orbifolds}
\vspace{1cm}}
\author{
Koushik Ray
\thanks{I.N.F.N.~~Fellow.}~\footnote{E-mail:~koushik@roma2.infn.it}\\
\small\em Dipartimento di Fisica\\
\small\em Universit{\`a} di Roma \ ``Tor Vergata'' \\
\small\em I.N.F.N.\  ---\ Sezione di Roma \ ``Tor Vergata'' \\
\small\em Via della Ricerca Scientifica, 1 \ \
\small\em 00133 \ Roma \ \ ITALY
}
\date{}
\maketitle
\begin{abstract}
\noindent We study issues pertaining to the 
Ricci-flatness 
of metrics on orbifolds resolved by D-branes. We find 
a K\"ahler metric on the three-dimensional 
orbifold $\C^3/\Z_3$, resolved by D-branes, 
following an approach due to Guillemin. This metric is
not Ricci-flat for any finite value of the 
blow-up parameter. Conditions for  
the envisaged Ricci-flat metric for finite values of 
the blow-up parameter are formulated in terms of a 
correction to the K\"ahler potential. 
This leads to an explicit construction of 
a Ricci-flat K\"ahler 
metric on the resolved orbifold. The correction 
constitutes a part of the superspace-interaction 
in the corresponding gauged linear sigma-model. 
\end{abstract}
\thispagestyle{empty}
\vfill\newpage
Duality symmetries in Type--I and Type--II string theories
necessitate incorporating D-branes in these theories, even 
though the branes may be of dimensions other 
than unity, and are not ``strings" in general. Apart from the fact
that these objects fit nicely into the conformal field theoretic 
framework, it looms large that D-branes will play a 
crucial role in our final understanding of space-time, as they can 
be used to probe the topology and geometry of space-time. 
In fact, D0-branes are deemed to be more efficient probes 
of space-time than strings in that they probe distance scales 
shorter than the string scale \cite{kabat}. These considerations 
have made D-branes interesting objects to study in recent times.

The world-volume theory of $n$ parallel D$p$-branes is 
described by a supersymmetric gauge theory, namely the ten-dimensional 
supersymmetric Yang-Mills theory, with gauge group $U(n)$, 
dimensionally reduced to $(p+1)$ dimensions \cite{bound}. 
The moduli space of this reduced theory has been interpreted
as the space-time sensed by D-branes \cite{bound}. In this sense 
space-time is a derived 
concept in the theory of D-branes. It has been found that 
it is possible to realize the moduli space of D-branes as
orbifolds by properly truncating the reduced theory to sectors
invariant under the action of some discrete group.
The theory of D-branes on two(complex)-dimensional 
ALE-spaces has been studied in detail \cite{dm,jm}. Such studies
realize Kr\"onheimer's construction of ALE-spaces in physical 
terms for the Kleinian subgroups of $SU(2)$ acting on the 
two-dimensional complex space, $\C^2$.
Fundamental strings on ALE-spaces have been considered earlier 
\cite{roma2}.

More recently, it has been found that D-branes can also be used to 
resolve three-dimensional orbifold singularities.
One can thus study the short-distance behavior of \cy manifolds 
as sensed by a D-brane probe in the neighborhood of a
resolved orbifold singularity. In order to study the theory 
of a single D-brane on $\C^3/\Gamma$, where 
$\Gamma$ is a discrete group of order $|\Gamma|$, 
one starts with $|\Gamma |$ number of D-branes on the covering space
$\C^3$, arranged 
in the regular representation of $\Gamma$. The world-volume 
theory of $|\Gamma |$ D-branes on $\C^3$ is a
four-dimensional gauge theory with $N=4$ supersymmetry and gauge group 
$U(|\Gamma |)$. The complex positions of the D-branes on
$\C^3$ are given by the scalars in the theory. 
Restricting to the sector of the D-brane Lagrangian invariant under 
the discrete group yields a gauged linear sigma-model. 
The nexus between the two-dimensional gauged linear sigma-model 
and the world-volume theory of a D1-brane probe has been exploited 
earlier \cite{bwb,gauge}. The moduli space of this linear sigma-model
is interpreted as the sub-stringy space-time.
This moduli space can be thought of as an internal space on which
D-branes move as points, with the world-volume of the D-brane 
lying in directions transverse to this internal space.
Now, if the discrete group $\Gamma$ is chosen to be a subgroup of $SU(3)$, 
then the resulting theory after truncation retains $N=1$ supersymmetry.
This construction has been realized for different choices of the discrete
group $\Gamma$; for example, 
$\Gamma=\Z_n$ for $n=3$, 5, 7, 9, 11 \cite{dgm,muto} and 
$\Gamma=\Z_2\times\Z_2$ \cite{brg,subir}. 
It should be mentioned that in these considerations one may start
with an arbitrary number $|\Gamma |$ of D-branes on the covering
space because the space
one considers is non-compact. The moduli space of the D-brane can
be viewed as a local description of a Calabi-Yau manifold 
only near an orbifold point.
Further generalization of these ideas to four dimensions has also
been considered \cite{mohri}.
It has been found that toric geometry provides a convenient 
language for such considerations \cite{dgm,leu,gre}. 
In the present paper we shall concentrate on the example of $\C^3/\Z_3$
\cite{dgm}.
As mentioned above, the theory of a D-brane on $\C^3/\Z_3$ is a 
four-dimensional $N=1$ theory, 
which is equivalent to the two-dimensional $N=2$ gauged linear
sigma-model \cite{wit2,asgr} upon dimensional reduction. 

Apart from the topological properties, some geometric, alias metric, 
properties of these resolved orbifolds have also been considered
\cite{dg}. It has been found that, starting from the flat K\"ahler 
potential on the covering space $\C^3$, and then adding 
the \FI terms, one can derive, by proper
gauging, a K\"ahler metric on the three-dimensional resolved
orbifold $\C^3/\Z_3$. However, this metric turns out to be a 
non-Ricci-flat one \cite{dg,curved}. 
It has also been found that the same prescription leads 
to the Eguchi-Hanson metric on the two-dimensional resolved
orbifold $\C^2/\Z_2$, which is Ricci-flat.
This dichotomy has been attributed to the fact that, unlike a resolved
$\C^2/\Z_2$,  the resolved orbifold $\C^3/\Z_3$, although K\"ahler, 
is not hyper-K\"ahler. In particular, it has been shown that
the metric derived in this fashion for $\C^3/\Z_3$ resolved into 
${\mathcal{O}}_{\C\p^2}(-3)$ assumes the form discovered by Calabi,
where ${\mathcal{O}}_{\C\p^2}(-3)$ is looked upon as having a 
canonical fiber over a base manifold of constant curvature.
This metric is Ricci-flat only for certain complex values of the 
\FI parameters, which is forbidden both physically and mathematically.

In this paper we shall follow an alternative 
approach to deriving K\"ahler metrics on toric 
varieties \cite{gui}. A novel feature
of this approach is that it makes it possible to derive the metric 
starting
from the toric data alone. This approach ensues from endowing a toric
variety $X$ with a symplectic structure arising from
a Hamiltonian action of the torus $\T_{\R}$ on $X$
\cite{cox}. It identifies the polytope $\Delta$ 
corresponding to the toric variety --- with $\Delta$ 
required to satisfy certain non-singularity 
and non-degeneracy conditions
(in particular, the polytope $\Delta$ must be Delzant \cite{gui,abreu})
--- with the moment polytope of the moment map of the Hamiltonian 
action of the torus on $X$ by a homeomorphism $X/\T_{\R}\rt\Delta$,
up to translation. This correspondence  provides a prescription for 
obtaining a K\"ahler metric on the variety \cite{gui}. We shall refer 
to this metric as the canonical metric in the sequel. 
This metric is not Ricci-flat in general.
However, starting from the canonical metric, 
one may go over to any other K\"ahler metric within the same 
K\"ahler class by adding a well-behaved function
to the (Legendre) dual
(to be defined below) of the K\"ahler potential \cite{abreu}. We 
pursue this line of approach here. 
We first find out the canonical metric, following \cite{gui}, on
the resolved orbifold $\C^3/\Z_3$, starting from the toric data as 
obtained in \cite{dgm}. We find that this metric is not 
Ricci-flat for any
finite value of the blow-up parameter. We then add a 
function $f$ to the (Legendre) dual K\"ahler potential and 
find out a differential equation for $f$, demanding 
that the new metric thus obtained is Ricci-flat. 
The solution of this equation leads to a Ricci-flat metric on the
orbifold $\C^3/\Z_3$ resolved by a D-brane.
Moreover, this additional function $f$ constitutes a part of 
the superspace-interaction in the linear sigma-model.     

In order to fix notations, let us start with some relevant
features of the present construction \cite{gui}.
Let $(X,\omega)$ be an effective Hamiltonian $\T^n$-space 
with a K\"ahler form $\omega$, and 
let $\varphi: X\rt \R^n$ denote the corresponding moment map 
associated to the Hamiltonian action of $\T^n$ on $X$. Let 
$\Delta$ denote the image of $X$ on $\R^n$ under the moment map, 
\ie $\Delta = \varphi(X)\subset\R^n$. $\Delta$ is referred to as the 
{\em moment polytope} and $X_{\Delta} = X$ is 
the toric variety associated to $\Delta$. 
Conversely, one can associate a toric variety $X_{\Delta}$ with
the above properties, to a Delzant polytope $\Delta$ 
in $\R^d$, such that $\Delta$ is the moment polytope of $X_{\Delta}$. 
Let us recall that the polytope $\Delta$ in $\R^d$ is called 
{\em Delzant} if there are $d$ edges meeting at each vertex $p$ of 
$\Delta$ and any edge meeting at $p$ can be given the form
$p+sv_i$, for $0\leq s\leq\infty$, where
$\{v_i\}$ is a basis of $\Z^d$ \cite{gui,abreu}.
The moment 
polytope $\Delta$ can be described by a set of inequalities of the form 
$\langle y, u_i\rangle \geq \lambda_i$, $i=0,2,\cdots d-1$. Here $u_i$ 
denotes the inward-pointing normal to the $i$-th $(n-1)$-dimensional 
face of $\Delta$
and is a primitive element of the lattice $\Z^n\subset \R^n$;
$\langle~,~\rangle$ denotes the standard scalar product in 
$\R^n$ and $y$ is an $n$-dimensional real vector.
We can thus define a set of linear maps, 
$\ell_i : \R^n\rt \R$, 
\begin{eqnarray}\label{li}
\ell_i(y) = \langle y, u_i\rangle - \lambda_i, \quad i=0,\cdots d-1. 
\end{eqnarray}
Denoting the interior of $\Delta$ by $\Delta^{\circ}$, 
$y\in\Delta^{\circ}$, if and only if $\ell_i(y) > 0$ for all $i$.

On the open $\T^n_{\C}$-orbit in $X_{\Delta}$, 
associated to a Delzant polytope $\Delta$, 
the K\"ahler form $\omega$ can be written in terms of a potential 
${{\mathcal{F}}}$ as \cite{gui}:
\begin{eqnarray}
\omega = 2i\pa\bar{\pa}{\mathcal{F}}, 
\end{eqnarray}
with
\begin{eqnarray}
{\mathcal{F}}=\frac{1}{2}~\varphi^{\star}
\left( \sum_{i=0}^{d-1}\lambda_i\ln\ell_i
+ \ell_{\infty} \right),
\end{eqnarray}
where $\varphi^{\star}$ denotes the pull-back of the moment 
map on $\Delta$ and we have defined 
\begin{eqnarray}
\ell_{\infty} = \sum_{i=0}^{d-1}\langle y, u_i\rangle .
\end{eqnarray}
The K\"ahler form $\omega$ can be written as:
\begin{eqnarray}
\omega = \frac{i}{2} \sum_{j,k=0}^{n-1} 
\frac{\pa^2{\mathcal{F}}}{\pa x_j\pa x_k} dz_j\wedge d\bar{z}_k.
\end{eqnarray}
The restriction to $\R^n = \re~\C^n$ of the K\"ahler metric 
corresponding to $\omega$ is the Riemannian metric given by, 
\begin{eqnarray}\label{remet}
ds^2=\sum_{j,k=0}^{n-1}\frac{\pa^2{\mathcal{F}}}{\pa x_j\pa x_k} 
dx_jdx_k.
\end{eqnarray}
Now, under the Legendre transform determined by the moment map 
$\varphi$, 
\begin{eqnarray}\label{legendre}
y_i = \frac{\pa{\mathcal{F}}}{\pa x_i}, 
\end{eqnarray}
the metric given by \eq{remet} is the pull-back of the metric
given by,
\begin{eqnarray}
ds^2=
\sum_{j,k=0}^{n-1}\frac{\pa^2{\mathcal{G}}}{\pa y_j\pa y_k} dy_jdy_k
\end{eqnarray}
on $\Delta^{\circ}$, where ${\mathcal{G}}$ is given by:
\begin{eqnarray}
{\mathcal{G}} = \frac{1}{2}\sum_{k=0}^{d-1}\ell_k(y)\ln\ell_k(y).
\end{eqnarray}
The inverse of the Legendre transform \eq{legendre} is:
\begin{eqnarray}\label{invleg}
x_i = \frac{\pa{\mathcal{G}}}{\pa y_i} + r_i, \quad i=0,\cdots , d-1,
\end{eqnarray}
where $r_i$ are constants. This means that up to a linear term in 
the co-ordinates
$y_i$, ${\mathcal{G}}$ is the K\"ahler potential Legendre-dual to 
${\mathcal{F}}$. Moreover, the matrix
\begin{eqnarray}
{\mathcal{G}}_{ij} = \frac{\pa^2{\mathcal{G}}}{\pa y_i\pa y_j},
\end{eqnarray}
evaluated at \eq{legendre}, \ie at
$y_i=\frac{\pa{\mathcal{F}}}{\pa x_i}$, is the inverse of the 
matrix 
\begin{eqnarray}\label{metric-f}
{\mathcal{F}}_{ij} = \frac{\pa^2{\mathcal{F}}}{\pa x_i\pa x_j}.
\end{eqnarray}

The Ricci-tensor for the metric \eq{metric-f} takes the following form:
\begin{eqnarray}\label{Ric-1}
R_{ij} &=& -\frac{1}{2}\frac{\pa^2\ln\det~\F}{\pa x_i\pa x_j}\\
\label{Ric-tensor}
&=& -\frac{1}{2}\sum_{k,l=0}^{n-1}
{\mathcal{G}}^{lj} \frac{\pa^2 {\mathcal{G}}^{ik}}{\pa y_k\pa y_l},
\end{eqnarray}
where ${\mathcal{G}}^{ij}$ denotes the inverse
of ${\mathcal{G}}_{ij}$.
Note that, $\F$ in \eq{Ric-1} refers to the matrix $\F_{ij}$, and not
the K\"ahler potential, unlike other places in the paper.
The Ricci-scalar for this metric is then derived by multiplying
\eq{Ric-tensor} with ${\mathcal{G}}_{ij}$, which is the inverse of 
the metric $\F_{ij}$ in the $y$ co-ordinates and is 
given by \cite{abreu},
\begin{eqnarray}\label{curv1}
R &=& -\frac{1}{2}\sum_{i,j=0}^{n-1}{\mathcal{F}}^{ij} 
\frac{\pa^2 \ln\det{\mathcal{F}}}{\pa x_i\pa x_j} \\\label{curvature}
&=& -\frac{1}{2}\sum_{i,j=0}^{n-1}\frac{\pa^2{\mathcal{G}}^{ij}}
{\pa y_i\pa y_j},
\end{eqnarray}
where ${\mathcal{F}}^{ij}$ denotes the inverse 
of ${\mathcal{F}}_{ij}$.
For our purposes, it will be convenient to use the matrix 
${\mathcal{G}}_{ij}$
in the coordinates $y$. One can, in principle, rewrite all
the relevant expressions in terms of the coordinates $x_i$ 
and the matrix ${\mathcal{F}}_{ij}$. 

With this introduction, let us now begin the study of metric properties
of the orbifold $\C^3/\Z_3$ resolved by D-branes. The description
of this orbifold along with its topological properties and phase structure
has been studied extensively \cite{dgm,brg}. We shall not repeat
the construction 
here. Let us start from the toric data obtained in \cite{dgm} for the 
resolved orbifold $\C^3/\Z_3$. The toric data is specified by
the charges of the components of the chiral superfields in terms of a 
$3\times 4$ matrix $\tilde{T}$ as,
\begin{eqnarray}\label{data} 
\tilde{T} =
\begin{pmatrix}
-1&1&0&0 \\ 
-1&0&1&0 \\
3&0&0&1
\end{pmatrix}.
\end{eqnarray}
Triangulating on all points yields the blow-up of 
the orbifold $\C^3/Z_3$. Given
the toric data \eq{data}, one can find out the inequalities 
describing the corresponding Delzant polytope \cite{cox}.
For this let us consider the fan in $\R^3$, corresponding to 
the toric data \eq{data} whose one-dimensional cone-generators 
are given by the four 
three-vectors constituting the four columns of $\tilde{T}$:
\begin{eqnarray}
\goth{e}_0 =
\begin{pmatrix}
-1\\-1\\3
\end{pmatrix},\quad
\goth{e}_1 =
\begin{pmatrix}
1\\0\\0
\end{pmatrix}, \quad
\goth{e}_2 =
\begin{pmatrix}
0\\1\\0
\end{pmatrix}, \quad
\goth{e}_3 =
\begin{pmatrix}
0\\0\\1
\end{pmatrix}.
\end{eqnarray}
Thus, this case corresponds to $n=3$ and $d=4$ of the above 
discussion. 
The Kernel of the matrix $\tilde{T}$ is 
\begin{eqnarray}
\ker{\tilde{T}} = \begin{pmatrix}
1\\1\\1\\-3
\end{pmatrix}.
\end{eqnarray}
Hence, the K\"ahler cone of the variety $X_{\Delta}$
associated to $\tilde{T}$ is given by 
\begin{eqnarray}\label{kcone}
\xi\equiv a_0 + a_1 + a_2 - 3a_3 > 0,
\end{eqnarray}
where we have denoted the image of the support function \cite{cox,gre} 
at $\goth{e}_i$ as $-a_i$ for $i=0,\cdots , 3$.
The moment map is given by 
\begin{eqnarray}\label{mommap}
|z_0|^2 + |z_1|^2 + |z_2|^2 - 3|z_3|^3 = \xi, 
\end{eqnarray}
which corresponds to the D-flatness condition of a $U(1)$ gauged 
linear sigma-model with four chiral superfields 
\cite{dgm,muto,wit2,asgr}. In \eq{mommap}, $z_i$,
$i=0, \cdots , 3$, denote the homogeneous coordinates on the toric 
variety. Let us also note that
this correspondence allows us to relate the blow-up parameter
$\xi$, which is proportional to the K\"ahler class of the variety, to
the \FI parameters of the linear sigma-model, using the charge
matrix \eq{data}, and hence study the phase structure \cite{dgm}. 
For all possible real values of the 
\FI parameters, $\xi$ remains positive in concordance with \eq{kcone}.
Physically, this means that the D-branes abstain from entering the
non-geometric phases of the linear sigma-model \cite{dgm,wit1}. 

Using the one-dimensional cone generators $\goth{e}_i$, 
we can now write down the Delzant polytope $\Delta$ in the {\em shifted} 
coordinates as the one defined by the inequalities:
\begin{eqnarray}
\xi - y_1 -y_2 + 3y_3 \geq 0, \\
y_1 \geq 0, \\
y_2 \geq 0, \\
y_3 \geq 0, 
\end{eqnarray}
where $\xi = a_0 + a_1 + a_2 - 3a_3$ is positive as above,
and we have applied
a shift of $a_i$ to each co-ordinate $y_i$, for $i=1,$ 2, 3. 
The above expressions 
are required to be strictly positive in the interior 
$\Delta^{\circ}$ of $\Delta$ and equality holds in each
of the above relations on the boundary of  the polytope defined by 
two-dimensional faces. We thus have the following linear
functions $\ell_i$, for $i=0,1,2,3$, 
\begin{eqnarray}\label{delpol}
\ell_1(y) = y_1, \quad
\ell_2(y) = y_2, \quad
\ell_3(y) = y_3, \quad
\ell_0(y) = \xi - y_1 -y_2 + 3y_3, \quad
\end{eqnarray}
whose positivity determine the Delzant polytope $\Delta$ up to shifts,
corresponding to the resolved orbifold $\C^3/\Z_3$.

We can now write down the potential ${\mathcal{G}}$ from \eq{delpol}:
\begin{eqnarray}\label{gorig}
{\mathcal{G}} = \frac{1}{2}\left[ y_1\ln y_1 + y_2\ln y_2 + y_3\ln y_3
+ (\xi - y_1 -y_2 +3y_3)\ln (\xi - y_1 -y_2 +3y_3) \right].
\end{eqnarray}
The inverse Legendre transform \eq{invleg} expresses 
the co-ordinates $x_i$ in terms of the co-ordinates $y_i$. 
The relations between these dual co-ordinates
can be written in the following form:
\begin{eqnarray}
\label{xy1}
(\xi + 3y_3)^3 y_3 &=& (1 + e^{2x_1} + e^{2x_2})^3 e^{2x_3}, \\
\label{xy2}
\frac{y_1}{y_2} &=& e^{2(x_1-x_2)}, \\
\label{xy3}
\frac{y_1}{A} &=& e^{2x_1}, 
\end{eqnarray}
where we have defined, 
\begin{eqnarray}
A = \xi -y_1 -y_2 + 3y_3,
\end{eqnarray}
and chosen the constants $r_i$ as $r_1=r_2=0$ and $r_3=2$ for simplicity.
Equations \eqs{xy1}{xy3} can be solved explicitly to write down 
expressions for $x_1$, $x_2$
and $x_3$ in terms of $y_1$, $y_2$ and $y_3$.
One can now calculate the matrix ${\mathcal{G}}_{ij} = 
\frac{\pa^2{\mathcal{G}}
}{\pa y_i\pa y_j}$, using the potential 
${\mathcal{G}}$ obtained in \eq{gorig}. It takes the form: 
\begin{eqnarray}
{\mathcal{G}}_{ij} = \frac{1}{2A}\begin{pmatrix}
1+A/y_1&1&-3\\
1&1+A/y_2& -3\\
-3&-3&9+A/y_3
\end{pmatrix}. 
\end{eqnarray}
Inverting this matrix  to write ${\mathcal{G}}^{ij}$, 
and then differentiating twice with respect to the 
co-ordinates, we can compute the Ricci-scalar according to
\eq{curvature} as:
\begin{eqnarray}\label{curv:old}
R = \frac{12(\xi^2 + 72y_3^2)}{(\xi+12y_3)^3}.
\end{eqnarray}
Thus, the metric on the variety is not Ricci-flat for any finite 
value of
the parameter $\xi$. However, the Ricci-scalar \eq{curv:old} vanishes 
as $1/\xi$ as $\xi$ is taken to infinity. 
The metric in this limit is given by 
\begin{eqnarray}\label{undeformed}
{\mathcal{G}}^{ij} = \begin{pmatrix}
2y_1 & 0&0\\
0&2y_2&0 \\ 0&0&2y_3
\end{pmatrix}.
\end{eqnarray}
The Ricci-tensor \eq{Ric-tensor} also vanishes for the metric 
\eq{undeformed}. Thus, in the limit of infinite $\xi$, we 
have a Ricci-flat metric. 

However, let us note that using the formulation of \cite{gui},
one can explicitly calculate the first Chern class of the variety
under consideration and check that it vanishes. For this, let us 
note that the K\"ahler class of the canonical metric can be 
expressed in terms of the parameters $\lambda_i$ as \cite{gui}:
\begin{eqnarray}
\frac{[\omega]}{2\pi} = -\sum_{i=0}^{d-1} \lambda_i\alpha_i,
\end{eqnarray}
where $\alpha_i$, $i=0,\cdots , {d-1}$ are constants. One can 
express the first Chern class of the variety $X_{\Delta}$ as a 
sum of the constants $\alpha_i$:
\begin{eqnarray}\label{c1}
\goth{c_1}(X_{\Delta}) = \sum_{i=0}^{d-1}\alpha_i.
\end{eqnarray} 
Physically, equation \eq{c1} expresses the first Chern class of the 
variety as a sum of the charges of the fields in the corresponding 
linear sigma-model.
Using the expression \eq{kcone} for $\xi$, which is proportional 
to the K\"ahler class of the variety, 
and noting that $\lambda_i = -a_i$, when the polytope $\Delta$ is 
written in terms of the {\em unshifted} co-ordinates, we find that 
\begin{eqnarray}
\alpha_0= \alpha_1 =\alpha_2 =1, \quad \alpha_3 = -3,
\end{eqnarray}
up to a constant of proportionality multiplying each of $\alpha_i$.
Hence, by \eq{c1}, the first Chern class of the resolved 
orbifold $\C^3/Z_3$ vanishes. This is
not surprising in view of the fact that the action of $\Z_3$ on
the chiral superfields were chosen in the beginning 
in such a way that the $\Z_3$ was a subgroup
of $SU(3)$ \cite{dgm}. Nonetheless, the present approach provides 
an explicit reconfirmation of the vanishing of the first 
Chern class of the resolved orbifold $\C^3/\Z_3$.

Let us now go over to explore the possibility of obtaining 
a Ricci-flat metric for finite values 
of the blow-up parameter $\xi$. For this purpose, let us note
\cite{abreu} that one can add a
function, which is smooth on some open subset of $\R^n$ containing
$\Delta$, to the potential ${\mathcal{G}}$, such that the Hessian of
the new potential is positive definite on $\Delta^{\circ}$, 
and derive a new K\"ahler 
metric in the same K\"ahler class as the canonical one.
The variety $X_{\Delta}$ endowed with the two different K\"ahler forms 
are related by a $\T^n$-equivariant symplectomorphism by virtue
of the function $f$ being non-singular \cite[Remark~3.1]{abreu}.  
For example, the function one has to add to the potential 
${\mathcal{G}}$, in order to derive the extremal K\"ahler 
metric on $\C\p^2\#\C\p^2$ in Calabi's form, was determined
utilizing this observation \cite{abreu}. 
Following this approach, let us as add
a function $f$ to ${\mathcal{G}}$, and define a new potential 
$\tilde{{\mathcal{G}}} = {\mathcal{G}} + \frac{1}{2}f$. 
Now the matrix $\G_{ij}$ corresponding to the 
potential $\tilde{{\mathcal{G}}}$ assumes the form 
\begin{eqnarray}
\tilde{{\mathcal{G}}}_{ij} = {\mathcal{G}}_{ij} + 
\frac{1}{2}\frac{\pa^2f}{\pa y_i\pa y_j}. 
\end{eqnarray}
One can then find out the K\"ahler metric by inverting 
$\tilde{{\mathcal{G}}}_{ij}$ and hence the curvature for this new 
metric. This will also give rise to a new $\tilde{{\mathcal{F}}}$ 
corresponding to
${\mathcal{F}}$ and also new co-ordinates $\tilde{x}$. What we 
propose to do next is to write down the general form of the metric 
for a function $f$ and then determine the function by demanding 
that the Ricci-tensor given by the formula \eq{Ric-tensor} vanishes. 
This gives a differential equation for
the function $f$. However, for practical purposes of obtaining a 
tractable equation for $f$, one needs to start with an ans\"atz 
for the function $f$. In view of the fact that the Ricci-scalar
\eq{curv:old} is a function
of $y_3$ alone, not depending on $y_1$ or $y_2$, let us assume that 
$f$ is a function of $y_3$ only, \ie $f=f(y_3)$. 
The matrix $\tilde{{\mathcal{G}}}_{ij}$ is then given by 
\begin{eqnarray}\label{deformed}
\tilde{{\mathcal{G}}}_{ij} = \frac{1}{2A}\begin{pmatrix}
1+A/y_1&1&-3\\
1&1+A/y_2& -3\\
-3&-3&9+A/y_3+Af''
\end{pmatrix},
\end{eqnarray}
where $A = \xi -y_1 -y_2 + 3y_3$, as before
and a prime denotes a differentiation with respect to $y_3$. 
The inverse of the matrix $\tilde{{\mathcal{G}}}_{ij}$ 
is given by 
\begin{equation} \label{invG}
\G^{ij}= 
\begin{pmatrix}
2y_1 - 2\frac{y_1^2}{F}\phi &
-2\frac{y_1y_2}{F}\phi &
6\frac{y_1y_3}{F} \\
-2\frac{y_1y_2}{F}\phi &
2y_2 - 2\frac{y_2^2}{F}\phi &
6\frac{y_2y_3}{F} \\
6\frac{y_1y_3}{F} &
6\frac{y_2y_3}{F} &
2\frac{y_3 (\xi+3y_3)}{F}
\end{pmatrix},
\end{equation}
where we have defined 
\begin{eqnarray}\label{Fphi}
F = \xi + 12 y_3 + (\xi + 3y_3)y_3f''  \qquad {\mathrm and}\qquad
\phi = \frac{F-9y_3}{\xi + 3y_3}.
\end{eqnarray}
The determinant of the matrix $\G_{ij}$ is  $\det \G_{ij} = 
\frac{F}{8Ay_1y_2y_3}$. Hence, throughout this paper we shall assume
$F$ to be nowhere-vanishing in order to keep $\G_{ij}$ non-singular.
For the metric \eq{invG}, the surviving terms in the 
different components of $R_{ij}$ as evaluated from 
\eq{Ric-tensor} take the following form:
\begin{eqnarray}
\label{r11}
-2R_{11} &=& 
\G^{11}\left(
\frac{\pa^2\G^{11}}{\pa y_1\pa y_1} +
\frac{\pa^2\G^{12}}{\pa y_1\pa y_2} +
\frac{\pa^2\G^{13}}{\pa y_1\pa y_3} 
\right) + 
\G^{13}\left(
\frac{\pa^2\G^{11}}{\pa y_1\pa y_3} +
\frac{\pa^2\G^{12}}{\pa y_2\pa y_3} +
\frac{\pa^2\G^{13}}{\pa y_3\pa y_3} 
\right), \\ 
\label{r12}
-2R_{12} &=& 
\G^{12}\left(
\frac{\pa^2\G^{11}}{\pa y_1\pa y_1} +
\frac{\pa^2\G^{12}}{\pa y_1\pa y_2} +
\frac{\pa^2\G^{13}}{\pa y_1\pa y_3} 
\right) + 
\G^{23}\left(
\frac{\pa^2\G^{11}}{\pa y_1\pa y_3} +
\frac{\pa^2\G^{12}}{\pa y_2\pa y_3} +
\frac{\pa^2\G^{13}}{\pa y_3\pa y_3} 
\right), \\ 
\label{r13}
-2R_{13} &=& 
\G^{13}\left(
\frac{\pa^2\G^{11}}{\pa y_1\pa y_1} +
\frac{\pa^2\G^{12}}{\pa y_1\pa y_2} +
\frac{\pa^2\G^{13}}{\pa y_1\pa y_3} 
\right) + 
\G^{33}\left(
\frac{\pa^2\G^{11}}{\pa y_1\pa y_3} +
\frac{\pa^2\G^{12}}{\pa y_2\pa y_3} +
\frac{\pa^2\G^{13}}{\pa y_3\pa y_3} 
\right), \\ 
\label{r22}
-2R_{22} &=& 
\G^{22}\left(
\frac{\pa^2\G^{12}}{\pa y_1\pa y_2} +
\frac{\pa^2\G^{22}}{\pa y_2\pa y_2} +
\frac{\pa^2\G^{23}}{\pa y_2\pa y_3} 
\right) + 
\G^{23}\left(
\frac{\pa^2\G^{12}}{\pa y_1\pa y_3} +
\frac{\pa^2\G^{22}}{\pa y_2\pa y_3} +
\frac{\pa^2\G^{23}}{\pa y_3\pa y_3} 
\right), \\ 
\label{r23}
-2R_{23} &=& 
\G^{23}\left(
\frac{\pa^2\G^{12}}{\pa y_1\pa y_2} +
\frac{\pa^2\G^{22}}{\pa y_2\pa y_2} +
\frac{\pa^2\G^{23}}{\pa y_2\pa y_3} 
\right) + 
\G^{33}\left(
\frac{\pa^2\G^{12}}{\pa y_1\pa y_3} +
\frac{\pa^2\G^{22}}{\pa y_2\pa y_3} +
\frac{\pa^2\G^{23}}{\pa y_3\pa y_3} 
\right), \\ 
\label{r33}
-2R_{33} &=& 
\G^{33}\left(
\frac{\pa^2\G^{13}}{\pa y_1\pa y_3} +
\frac{\pa^2\G^{23}}{\pa y_2\pa y_3} +
\frac{\pa^2\G^{33}}{\pa y_3\pa y_3} 
\right). 
\end{eqnarray}
Thus, the Ricci-tensor vanishes if the following five 
equations are satisfied:
\begin{eqnarray}
\label{r111}
\frac{\pa^2\G^{11}}{\pa y_1\pa y_1} +
\frac{\pa^2\G^{12}}{\pa y_1\pa y_2} +
\frac{\pa^2\G^{13}}{\pa y_1\pa y_3} &=& 0,\\
\label{r112}
\frac{\pa^2\G^{11}}{\pa y_1\pa y_3} +
\frac{\pa^2\G^{12}}{\pa y_2\pa y_3} +
\frac{\pa^2\G^{13}}{\pa y_3\pa y_3} &=&0,\\
\label{r221}
\frac{\pa^2\G^{12}}{\pa y_1\pa y_2} +
\frac{\pa^2\G^{22}}{\pa y_2\pa y_2} +
\frac{\pa^2\G^{23}}{\pa y_2\pa y_3} &=& 0,\\
\label{r222}
\frac{\pa^2\G^{12}}{\pa y_1\pa y_3} +
\frac{\pa^2\G^{22}}{\pa y_2\pa y_3} +
\frac{\pa^2\G^{23}}{\pa y_3\pa y_3} &=& 0,\\
\label{r331}
\frac{\pa^2\G^{13}}{\pa y_1\pa y_3} +
\frac{\pa^2\G^{23}}{\pa y_2\pa y_3} +
\frac{\pa^2\G^{33}}{\pa y_3\pa y_3} &=& 0.
\end{eqnarray}
Let us note that adding the equations \eq{r111}, \eq{r221} and \eq{r331}
yields the equation of vanishing Ricci-scalar, as derived from 
\eq{curvature}.
Using the explicit form of the matrix \eq{invG}, one can now write 
down the equations
for $F$ corresponding to \eqs{r111}{r331}. The equations following from
\eq{r111} and \eq{r221} are identical; those from \eq{r112} and \eq{r222}
are also identical. We are thus left with three equations following 
from \eq{r111}, \eq{r112} and \eq{r331}. These are, respectively,
\begin{eqnarray}
\label{eqn1}
F - y_3 F' - \phi F &=& 0 \\
\label{eqn2}
2y_3(F')^2 - ({\phi}'F - {\phi}F')F - y_3FF'' - 2FF' &=& 0 \\
\label{eqn3}
2y_3(\xi+3y_3)(F')^2 - y_3(\xi+3y_3)FF'' - 2(\xi+9y_3)FF' 
+ 12F^2 &=& 0.
\end{eqnarray}
However, these three equations are not independent. 
Equations \eq{eqn2} and \eq{eqn3} can be solved by using only 
\eq{eqn1}. Thus, finally, we need to solve only \eq{eqn1}
in order to determine $F$, \ie the function $F$ is {\em not} 
over-determined by the condition of Ricci-flatness of the metric.
Rewriting \eq{eqn1} by plugging in the expression of $\phi$
from \eq{Fphi}, we derive the following equation, 
\begin{eqnarray}\label{eqnF}
y_3(\xi+3y_3)F'  + (F-\xi-12y_3)F = 0.
\end{eqnarray} As a check on consistency, let us note  that \eq{eqnF}
solves the equation for vanishing Ricci-scalar, which, by
\eq{curvature}, can be written as  
\begin{eqnarray}\label{non-lin}
y_3(\xi+3y_3)^2{F}{F}'' &+& 2(\xi+3y_3)(\xi+12y_3){F}{F}' \nonumber \\
&-&2y_3(\xi+3y_3)^2({F}')^2 +6 (F-3\xi -18 y_3){F}^2 = 0.
\end{eqnarray}
We shall linearize and solve \eq{eqnF} for $F$. 
Once $F$ is determined, we can use \eq{Fphi} to determine $f''$, 
and this yields the Ricci-flat metric, using $f''$ in \eq{invG}.
In order to solve  for the K\"ahler potential one has to solve 
for $f$. Let us point out that using the definition of 
$F$ from \eq{Fphi} in \eq{eqnF},
we have the following relation between $F$ and $f$:
\begin{eqnarray}
f'' = -F'/F.
\end{eqnarray}

In order to solve \eq{eqnF}, we divide both sides of \eq{eqnF} by $F$,
as $F$ is nowhere-vanishing, to derive the first order linear equation,
\begin{eqnarray}\label{chi}
y_3(\xi+3y_3){\chi}' + (\xi+12y_3)\chi =1,
\end{eqnarray}
where $\chi = 1/F$.
Equation \eq{chi} can be solved to derive the following expression 
for $F$:
\begin{eqnarray}
F = \frac{9y_3(\xi+3y_3)^3}{c+(\xi+3y_3)^3},
\end{eqnarray}
where $c$ is a constant of integration. This in turn yields the 
following expression for $f''$:
\begin{eqnarray}\label{fpp-sol}
f'' = \frac{9y_3(\xi+3y_3)^3 - (\xi+12y_3)\left(c + (\xi+3y_3)^3\right)
}{y_3(\xi+3y_3)\left(c+(\xi+3y_3)^3\right)}.
\end{eqnarray}
Using this in the expression \eq{invG} for $\G^{ij}$, we 
have a Ricci-flat metric on the variety.
Equation \eq{fpp-sol} can be integrated to solve for $f$ and hence
determine $\G$. Thus, we finally have the explicit form of a Ricci-flat 
metric and the K\"ahler potential on the variety in a neighborhood of 
the resolved orbifold singularity $\C^3/\Z_3$.

Finally, let us discuss the physical meaning of the function $f$.
Let us recall that it is customary in the context of gauged linear 
sigma-models to work with the metric induced from the standard metric 
on $\C\p^n$, in which one can embed the space described by
the D-flatness condition. 
This metric, although K\"ahler, is not Ricci-flat \cite{wit2}. 
It is believed that there exists a unique Ricci-flat K\"ahler 
metric for large values of the \FI parameter, $\xi$, in the present 
case. Moreover, this 
Ricci-flat metric differs from the one induced from 
the Fubini-Study metric on $\C\p^n$
in such a way that the difference between the cohomologous K\"ahler forms 
corresponding to the two metrics is given by $-i\pa\bar{\pa}T$. 
The difference between the corresponding the K\"ahler potentials, $T$, 
provides the superspace interaction term in the Lagrangian of the
linear sigma-model, namely,
\begin{eqnarray}\label{suint}
\int d^2xd^4\theta T,
\end{eqnarray}
in the notation of \cite{wit2}. In the limit of infinite coupling, 
the linear sigma-model \cite{wit2} differs from the conformal invariant 
non-linear sigma-model by the term \eq{suint}. Therefore, in order to
learn about the conformal invariant model from the linear one of 
\cite{wit2}, one needs to know the form of $T$ in \eq{suint}.

Now, let us recall the following results from \cite{gui}. 
It can be proved that the de Rham cohomology classes 
in $H^2(X,\R)$ of the canonical K\"ahler form $\omega$
obtained through Guillemin's construction and the metric induced on
$\C^3/\Z_3$ from the Fubini-Study metric on $\C\p^N$ are the same.
It can also be proved \cite{gui} that:
there exists a smooth function, $Q$, on $\R^n$, such that
\begin{eqnarray}
\omega_{\mathrm{FS}} = \omega + i\pa\bar{\pa}\varphi^{\star}Q.
\end{eqnarray}
Moreover, {\em $Q$ is unique up to an additive constant}. Here 
$\omega_{\mathrm{FS}}$ denotes the pull-back 
from the Fubini-Study metric on $\C\p^N$.
An explicit form for $Q$ is given in \cite{gui}. The 
K\"ahler classes of the forms corresponding to 
${\mathcal{G}}$ and $\tilde{{\mathcal{G}}}$ are the same 
\cite{abreu}. The K\"ahler form corresponding 
to $\tilde{{\mathcal{G}}}$ can be written as :
\begin{eqnarray}
\tilde{\omega} = \omega_{\mathrm{FS}} + i\pa\bar{\pa}(f-\varphi^{\star}Q ).
\end{eqnarray}
In other words, the expression $(f-\varphi^{\star}Q)$ is the 
difference between the potentials corresponding to the 
induced Fubini-Study metric on the resolved orbifold
and the Ricci-flat K\"ahler metric \eq{invG}. 
Hence, in the limit of strong coupling this may be identified with
${T}$ in \eq{suint}, providing the  ``irrelevant" superspace 
interaction term needed to drive the linear sigma-model to the 
conformal invariant limit \cite{wit2}. 

To conclude, in this paper we have followed a construction due to 
Guillemin to derive a K\"ahler metric on the orbifold $\C^3/\Z_3$
resolved by D-branes, in the neighborhood of the blown-up 
orbifold point. The metric derived following \cite{gui}
is not Ricci-flat for any finite value of the blow-up parameter.
Then, making use of the observation that one can add a well-behaved
function to the Legendre-dual K\"ahler potential ${\mathcal{G}}$
\cite{abreu}, we
determined a function $f$ which can be added to the canonical 
potential to obtain a Ricci-flat metric. In this approach one 
derives an explicit form of the Ricci-flat metric, for all 
values of the blow-up parameter. 
The novelty of the construction of \cite{gui} is that it allows 
to derive the metric on the variety and the corresponding 
K\"ahler potential starting from the toric data alone, 
and hence avoids the complications due to the choice of
seed metric \cite{dg}. The canonical metric itself can be thought
of as the seed metric in this context. 
Since the resolution of the orbifold  $\C^3/Z_3$
as effected by D-branes is encoded in the toric data $\tilde{T}$, 
which correspond to charges of the fields in the corresponding 
linear sigma-model \cite{dgm},
the metric \eq{invG} with $f''$ given by \eq{fpp-sol} is the metric 
on the resolved orbifold sensed by D-branes in a
neighborhood of the orbifold point. It will be interesting 
to check this result by direct world-sheet computations as 
suggested in \cite{dg}.

In view of the result of Guillemin relating the K\"ahler potentials
corresponding to the canonical metric and 
the metric induced on the resolved orbifold from the Fubini-Study metric 
on $\C\p^N$, by a function $Q$, we find that the function $f$, combined 
with $Q$, gives the superspace interaction term in the strong
coupling regime of the 
linear gauged sigma-model, which is required to drive the linear 
sigma model to the non-linear one with conformal symmetry.
It will be interesting to see how this formulation can be used 
to drive the linear sigma-model to the limit of strong coupling with
a large \FI parameter. It is expected that this 
formulation will shed some light on the emergence of conformal
symmetry in this context. 

It will be interesting to extend this formulation 
for other choices of ans\"atz for $f$.
Another interesting point to note is that 
while Calabi's form of the metric is derived from the condition that
the metric on the base of the resolution ${\mathcal{O}}_{\C\p^2}(-3)$
has a constant determinant, it is not so in the present case.
Here we demand the vanishing of the Ricci-tensor 
$R_{ij}$ as given by \eq{Ric-tensor}, 
which is different from demanding constant determinant on the base.
It will be interesting, at least mathematically,
to find out an explicit relation between the two metrics.
Finally, it should be possible, by following the same line of arguments
as here, to generalize this formulation for other cases, for example,
$\Z_n$ and $\Z_2\times\Z_2$.
We hope to return to some of these issues in near future.

\noindent{\bf Acknowledgements:~~}
I take this opportunity to thank M Bianchi and A Sagnotti 
for encouragement and enlightening discussions.
I also thank D Polyakov for useful discussions.

\end{document}